\documentclass[twocolumn,showpacs,preprintnumbers,amsmath,amssymb,superscriptaddress, aps]{revtex4}

\usepackage{graphicx}% Include figure files
\usepackage{dcolumn}% Align table columns on decimal point
\usepackage{bm}% bold math
\usepackage{color}

\begin{document}
\newcommand{\aver}[1]{\ensuremath{\big<#1 \big>}}
\newcommand{\cred}{\color{red}}

\title{Mixture of bosonic and spin-polarized fermionic atoms in an optical lattice}

\author{Lode Pollet}
%\email{pollet@itp.phys.ethz.ch}
\affiliation{Theoretische Physik, ETH Z\"urich, CH-8093 Z\"urich, Switzerland}

\author{Corinna Kollath}
\affiliation{Universit{\' e} de Gen{\` e}ve, 24 Quai Ernest-Ansermet, CH-1211 Gen{\` e}ve, Switzerland}

\author{Ulrich Schollw\"{o}ck}
\affiliation{Institute of Theoretical Physics C, RWTH Aachen University, D-52056 Aachen, Germany}

\author{Matthias Troyer}%
\affiliation{Theoretische Physik, ETH Z\"urich, CH-8093 Z\"urich, Switzerland}

\date{\today}

\begin{abstract}
We investigate the properties of trapped Bose-Fermi mixtures for experimentally
relevant parameters in one dimension.
%The effect of the attractive interaction between the fermions and the bosons
The effect of the attractive Bose-Fermi interaction 
onto the bosons is to
deepen the parabolic trapping potential, and to reduce the bosonic
%repulsion in higher order. This reduction would lead to an increase of the bosonic coherence at constant temperature.  The opposite effect was observed in the experimental ${}^{87}$Rb - ${}^{40}$K
repulsion in higher order,  leading to an increase in bosonic coherence.  The opposite effect was 
% observed in the experimental ${}^{87}$Rb - ${}^{40}$K systems, most likely due to a sharp rise in 
observed in  ${}^{87}$Rb - ${}^{40}$K experiments, most likely due to a sharp rise in 
temperature. 
%We discuss the features which could be observed experimentally if temperature remains low, such as
We also discuss low-temperature features, such as
a bosonic Mott insulator transition driven by the fermion concentration, and the formation of composite particles such as polarons and molecules.

\end{abstract}


\maketitle

\section{Introduction}

Systems of interacting bosons and fermions occur frequently in nature. Usually, the bosons act as carriers of force between the fermionic particles. In high energy physics, quarks exchange gluons via the strong force, while in solid state physics electrons can interact via light or lattice vibrations.
Most prominent examples of such systems are conventional BCS superconductivity (caused by an effective
attractive interaction between the fermions induced by the electron-phonon
coupling), the Peierls instability (a charge density wave) and the formation
of polarons, which in solids are electrons dressed by a cloud of
phonons. 
There are only a few condensed matter systems in which the influence of fermions
onto bosons has been investigated. One of them are mixtures of bosonic ${}^4$He and
fermionic ${}^3$He, in which a shift of the transition temperature between normal
and superfluid ${}^4$He as a function of ${}^3$He concentration was observed.

In the field of ultracold gases 
fermions and bosons are on an equal footing. 
The choice of different atomic species \cite{Truscott01, Schreck01,Modugno02,Silber05}, the use of Feshbach resonances
\cite{Hadzibabic03,Inouye04,Ospelkaus06_fesh} and optical lattice
potentials \cite{Gunter06,Ospelkaus06} give
almost unrestricted access to all parameters of these systems, offering the possibility to study the influence of the species
onto each other and to investigate open questions from other areas of
physics in a new context. 
Theoretical approaches~\cite{Buechler03,Cazalilla03,  Albus03, Kuklov03,Roth04,Cramer04, Mathey04, Lewenstein04, Imambekov05, Pollet06} have proposed a whole variety of quantum phases present in
homogeneous Bose-Fermi mixtures at low temperature, ranging from a
charge-density wave, over a fermionic pairing phase, to polaronic properties, and even to
phase separation. 

In recent
experiments two groups independently succeeded in
the stabilization of bosonic ${}^{87}$Rb and fermionic ${}^{40}$K  in a
three-dimensional optical lattice~\cite{Gunter06,Ospelkaus06}. They focused 
on the loss of bosonic phase
coherence and on the increase of the bosonic density by varying the fermionic concentration.
The trapping potential and the finite
temperature make the interpretation of the observed quantities however challenging. 

Here, we study the interplay of a trap, finite temperature and
strong interparticle interactions, which lead to physics quite different from the
homogeneous case. In particular, while the
addition of fermions induces a quantum phase transition from the Mott
insulating to the superfluid phase at larger bosonic repulsion strength in a homogeneous lattice,  the presence of a trapping
potential makes the situation more involved because of an extra, effective strongly-inhomogeneous trapping potential. Despite the large Bose-Fermi
coupling, it turns out that our results for the trapped, mixed system can be well
understood in terms of first and second order corrections to the bosonic
Hamiltonian. For low bosonic and fermionic densities, we illustrate that the formation of bound pairs invalidates the picture of a perturbational correction by the fermions on the bosons.

%%%%%%%%%%%%%%%%%%%%%%%%%%%%%%%%%%%%%%%%%%%%%%%%%%%%%%%%%%%%%%%%%%%%%%%%%%%%%%
\section{Model}

A mixture of bosonic and spin-polarized fermionic atoms in an optical lattice can be described by the Bose-Fermi Hubbard Hamiltonian,
\begin{eqnarray}
H & = & - \sum_{\langle i, j \rangle}^L\left(  J_{\rm B} \hat{b}^{\dagger}_i \hat{b}_j + J_{\rm F} \hat{c}^{\dagger}_i \hat{c}_j  + {\rm h.c.} \right) + \nonumber \\
{} & {} & \sum_i^L \frac{U_{\rm BB}}{2} \hat{n}_{{\rm B},i}(\hat{n}_{{\rm B},i}-1) + \sum_i^L U_{\rm BF} \hat{n}_{{\rm B},i} \hat{n}_{{\rm F},i} + \nonumber \\
{} & {} & \sum_i^L \epsilon_{{\rm B},i} \hat{n}_{{\rm B}, i} + \sum_i^L \epsilon_{{\rm F},i} \hat{n}_{{\rm F}, i},\label{eq:bfh}
\end{eqnarray}
where $\hat{c}^{\dagger}_i (\hat{b}^{\dagger}_i) $ and $\hat{c}_i (\hat{b}_i)$
are the corresponding creation and annihilation operators for the fermions
(bosons), and $\hat{n}_{{\rm X},i}$ is the number operator on site $i$ for
species X=B,F. The $J_{\rm X}$-terms  are the hopping, the $\epsilon_{\rm X}$-terms describe the external trapping potential, and $U_{\rm BB}$ and $U_{\rm BF}$ denote the on-site interaction strength between bosonic atoms and beween a fermionic and bosonic atom, respectively. 
The effective parameters of the Bose-Fermi Hubbard model are deduced from the
experimental parameters of Ref.~\cite{Ospelkaus06,Gunter06} using a
%tight-binding approximation~\cite{Jaksch98, JakschPhD, Buechler03}. Taking the scattering
tight-binding approximation~\cite{Jaksch98, Buechler03}. Taking the scattering
lengths as $a_{\rm BB}/a_0 = 102 \pm 6$~\cite{Buggle04} 
and $a_{\rm BF}/a_0 = -205 \pm 5$~\cite{Ferlaino06}, where $a_0$ is the Bohr
radius, we note that  $U_{\rm BF}/ U_{\rm BB} \approx -2$, a ratio which is
almost constant for all optical lattice depths. We took a wavelength $\lambda
= 1064$nm for the optical lattice potential, and frequencies $\omega_{\rm B} =
2\pi \cdot 30 {\rm Hz}$ and $\omega_{\rm F} =2 \pi \cdot  37 {\rm Hz}$ for the harmonic confinement. 
To determine the state of the mixture we use two numerically exact methods: at finite temperatures the canonical two-body Bose-Fermi worm Quantum Monte Carlo (QMC) algorithm~\cite{Pollet06}, and at zero temperature the density-matrix renormalization-group method (DMRG)~\cite{White1992}.
%%%%%%%%%%%%%%%%%%%%%%%%%%%%%%%%%%\paragraph{homogeneous system}

\begin{figure}[t]
%\centerline{\includegraphics[scale=0.35,angle=270]{fig_zero}}
\centerline{\includegraphics[angle=270, scale=0.35]{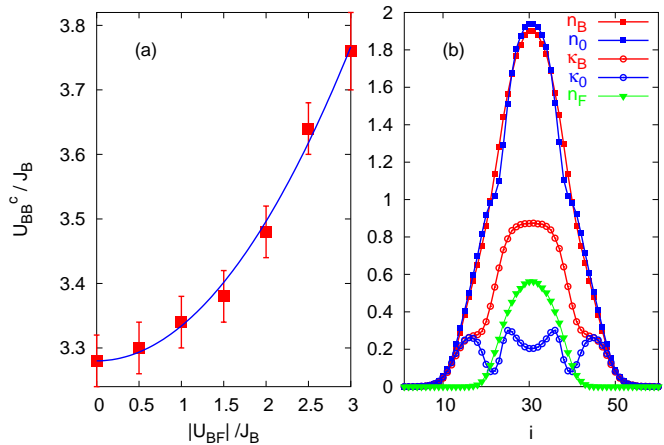}}
\caption{(Color online) (a) Shift in critical $U_{\rm
  BB}^{\rm c}$ of the Mott transition for attractive $U_{\rm
  BF}$. The system consists of $13$ fermions and $64$ bosons on a homogeneous lattice of $64$
  sites. Both species have unit hopping and the inverse temperature is $\beta
  = 64/J_{\rm B}$.  The critical value $U_{\rm BB}^{\rm c}/J_{\rm B} = 3.28 \pm 0.04$ at $U_{\rm
  BF}/J_{\rm B} = 0$ is taken from Ref.~\cite{PolletPhD}. At finite $U_{\rm BF}$ the
  transition is located where the Green function has the same algebraic decay
  as in the purely bosonic case. The solid curve is a parabolic
  fit. (b) Comparison between the calculation in the
  presence of fermions and their approximation 
  by a site-dependent potential for a trapped system of 60 sites, 40
  bosons, 8 fermions,  $\beta = 1/J_{\rm B}$ and optical
  potentials $V_0 = 6 E_{\rm R}$ ($U_{\rm BB}/J_{\rm B} =11.89$). Here,
  $E_{\rm R}= \hbar^2k^2/2m_{\textrm{Rb}}$ is the bosonic recoil
energy.  $n_{\rm
    B(F)}$ denotes the bosonic (fermionic) density for the mixture, $n_0$ is
  the density obtained in the approximation. Analogous for the density
  fluctuations $\kappa_{\rm B}$ and $\kappa_0$.  \label{fig:zero}
}
\end{figure} 

\section{Effects of fermions on bosons}

\subsection{Induced potentials and interactions}

In a mixture, the lowest order effect of the fermions
is a mean-field shift $U_{\rm BF} \aver{n_{\rm
    F,i}}$ of the potential experienced by the bosons  \cite{Buechler03, Buechler04}. In a homogeneous system with a fixed particle number the
shift in the potential has no consequences besides adding a constant to the energy. 
The next order effect is an induced attractive
interaction between the bosons \cite{Buechler03, Buechler04}. A similar effect
is well known from conventional superconductivity, where the phonons (bosons) induce an
effective electron-electron interaction. The induced interaction shows up most clearly in a shift of the
critical $U_{\rm BB}^c/J_{\rm B}$ of the bosonic superfluid-Mott transition
while varying the interspecies interaction as shown in  Fig.~\ref{fig:zero}(a). As expected
for an induced \emph{attractive} interaction, we find a shift to \emph{larger} values of $U_{\rm
  BB}/J_{\rm B}$. At small $U_{\rm BF}$ the shift is proportional to $U_{\rm BF}^2$ in agreement
with perturbative calculations~\cite{Buechler04}. Therefore the presence of
fermions can induce a phase transition from a Mott-insulating to a superfluid phase.
%%%%%%%%%%%%%%%%%%%%%%%%%%%

\begin{figure} [t]
%\centerline{\includegraphics[scale=0.35,angle=270]{fig_ubf}}
\centerline{\includegraphics[angle=270,scale=0.35]{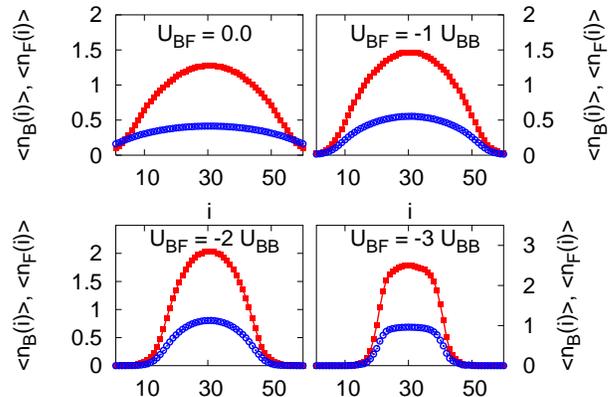}}
\caption{(Color online) Dependence of bosonic (red, upper curve) and fermionic
  density (blue, lower curve) profiles on the interspecies interaction strength.
 In the system there are 60 sites,  50 bosons, 20 fermions, optical potential is $V_0 = 3
 E_{\rm R}$ ($ U_{\rm BB} / J_{\rm B} = 4.26$), and the inverse temperature
 $\beta = 4.26 /J_{\rm B}$. 
}
\label{fig:ubf}
\end{figure}

In the following we discuss how a parabolic trap and finite temperature change this picture. 
%With a parabolic confining potential the shift in the potential becomes important~\cite{Buechler03}. 
To separate the effect of the effective trapping from the induced interaction,
we generate an effective site-dependent potential for the
bosons by replacing the Bose-Fermi interaction operator by the effective potential
$\mu_i \hat{n}_{{\rm B}, i} = U_{\rm BF} \langle n_{{\rm F},i} \rangle
\hat{n}_{{\rm B}, i}$. 
This deviates from the disordered chemical potential approach of Ref.~\cite{Ospelkaus06} and from a mean-field approximation:
The exact fermionic density distribution of the mixture serves as input for a second, purely bosonic simulation. In Fig.~\ref{fig:zero}(b), we compare the resulting bosonic
density and compressibility profiles. We observe that the density
profiles are quite well reproduced, confirming that the dominant effect of the
fermions is to modify the effective potential for the bosons. However, looking at higher order quantities such as the density
fluctuations $ \langle n_{{\rm B}, i}^2 \rangle - \langle n_{{\rm B},i}
\rangle^2$ 
 we find significant discrepancies. 
 %The role of
%the fermions thus goes beyond the creation of the effective chemical
%potential for the bosons. 
In particular, we see Mott plateaus in the
approximation (signaled by dips in the variance of the density in Fig.~\ref{fig:zero}(b)) that are absent in the full QMC simulation. Around these dips the difference between the two curves is around eighty percent. This is a clear
signature for a fermion-induced attractive Bose-Bose interaction, reducing the
bare repulsion $U_{\rm BB}$ (cf. Fig.~\ref{fig:zero}(a)).
We note that the visibilities, discussed below, are rather well reproduced in the approximation, indicating that the effective potential is the dominant effect as far as this experimental quantity is concerned (which is surprising seen the large values of $U_{\rm BF}$). 

Having gained an understanding of the relevant mechanisms, we proceed in section~\ref{subsec:sim} with the results of two simulations where we vary experimental control parameters, namely the Bose-Fermi coupling  and the fermionic concentration, and then compare our results to experiment in section~\ref{subsec:exp}.

\subsection{Results at low and constant temperature}\label{subsec:sim}

Fig.~\ref{fig:ubf} shows density profiles for different values of the
attractive interaction strength $U_{\rm BF}$. 
In the absence of a
boson-fermion interaction, all particles are smeared out over the
lattice. Turning the interspecies interaction on, we see in Fig.~\ref{fig:ubf}
that both species accumulate in the trap center. 
%The fermions are surrounded
%by a fluctuating number of bosons which can be viewed as a polaron, as was
%shown for the homogeneous, incommensurate case ~\cite{Mathey04}.
The fermions are pinned down
in the trap center (cf.~ Fig.~\ref{fig:ubf}), despite their light bare mass. They lower the effective potential in the center of the trap as
 $U_{\rm BF} \aver{n_{\rm F,i}}$, causing the accumulation of bosons.

\begin{figure}[ t]
%\centerline{\includegraphics[scale=0.35,angle=270]{fig_nf}}
\centerline{\includegraphics[angle=270, scale=0.35]{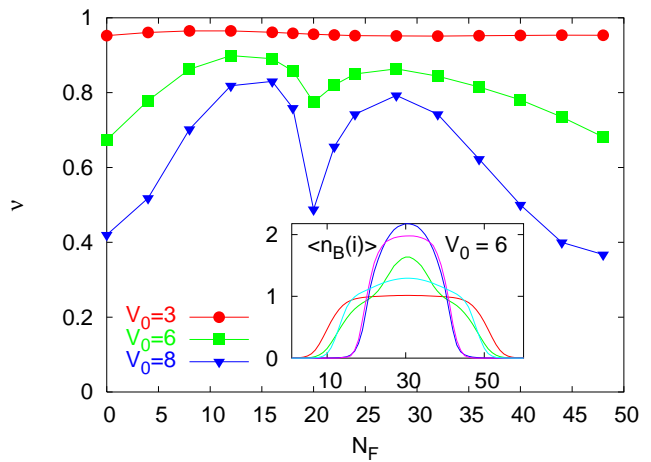}}
\caption{(Color online) The bosonic visibility $\nu$ as a function of fermionic number
  for a system of $L=60$ sites, $N_B=40$ bosons and inverse temperature $\beta = 1/J_{\rm
    B}$ for optical potentials $V_0 = 3, 6, 8 E_{\rm R}$ (or $U_{\rm
    BB}/J_{\rm B} = 4.26, 11.89, 21.58$, respectively). The inset shows the
  bosonic density profiles at $V_0 = 6 E_{\rm R}$ for $N_{\rm f} = 0, 40, 4, 20, 16$ bottom to top in the trap center. 
  }
\label{fig:concentration}
\end{figure}

The effect of an inhomogeneous effective potential can be
 seen even more clearly by varying the fermionic concentration instead of the interaction strength
 $U_{\rm BF}$.
In Fig.~\ref{fig:concentration} we show the dependence of the bosonic
 visibility on the number of fermions with a fixed number
of bosons, a setup similar to recent experiments~\cite{Ospelkaus06,Gunter06}.
The bosonic visibility is defined by $\nu = (\rho_{\rm B}
(0) - \rho_{\rm B} (\pi))/(\rho_{\rm B} (0) + \rho_{\rm B} (\pi))$ where
$\rho_{\rm B}(k)$ is the value of the bosonic momentum distribution at momentum
$k$. The visibility is often taken as a measure of the
 coherence of the bosons. 
 % Deep in the superfluid phase the
 % visibility is close to one due to the sharp peak of the momentum
 % distribution, while in the Mott phase the visibility is low due to the flat
 % momentum distribution (see however Ref.~\cite{JasonHo} for a different analysis).

For shallow lattice potentials, a slight increase in the visibility for an intermediate
number of fermions is seen, due to an increase in the bosonic density in the center of the trap caused by the fermions. 
%For moderate lattice potentials, on the other hand, 
For moderate lattice potentials
one observes a complex, non-monotonic behavior with large 
variations. If a few fermions are admixed to a bosonic system, 
the fermions -- spreading over several sites in
the center of the trap --  cause a strongly inhomogeneous
effective potential for the bosons. The effective potential exhibits a deep
minimum in the center of the trap and causes
the bosons to accumulate in this region.
If the purely bosonic system was superfluid 
(cf.~Fig.~\ref{fig:concentration}, $N_{\rm F}=0$, $V_0 = 3 E_{\rm R}$) the effective
potential causes a superfluid state with higher filling in the center of the
trap, {\it  slightly increasing} the bosonic visibility.
If the bosonic system exhibited a broad  $n_{\rm B} \approx 1$ Mott plateau (cf.~inset
Fig.~\ref{fig:concentration}, $N_{\rm F}=0$, $V_0 \ge 6 E_{\rm R}$), this plateaux is
partially destroyed resulting in a {\it large 
rise of the visibility}. 
The mechanism holds until there are enough
fermions present to form a band insulating region (cf.~Fig.~\ref{fig:concentration}, $N_{\rm F} \approx
14$). For such and higher fermion numbers, the effective potential induced by the fermions follows the curvature of the external trap  over the region occupied by the fermions with a sudden increase at its
boundaries. The number of fermions sets the length of an
effective system for the bosons, and controls the bosonic filling. In this
approximately parabolically trapped effective systems insulating regions can form
for $N_{\rm B}/N_{\rm F} = 40/20$ (as shown), but also for  $N_{\rm B}/N_{\rm F} \approx 60/30 \quad {\rm or} \quad 20/10$, yielding strong dips in the visibility~\cite{Batrouni02,
  KollathZwerger2004}.

%%%%%%%%%%%%%%%%%%%%%%%%%%%%%%%%%%%%%%%%%%%%%\paragraph

%%%%%%%%%%%%%%%%%%%%%%%%%%%%%%%%%%%%%%%%%%%%%%

\begin{figure}[ t]
%\centerline{\includegraphics[scale=0.35,angle=270]{fig_temp}}
\centerline{\includegraphics[angle=270,scale=0.35]{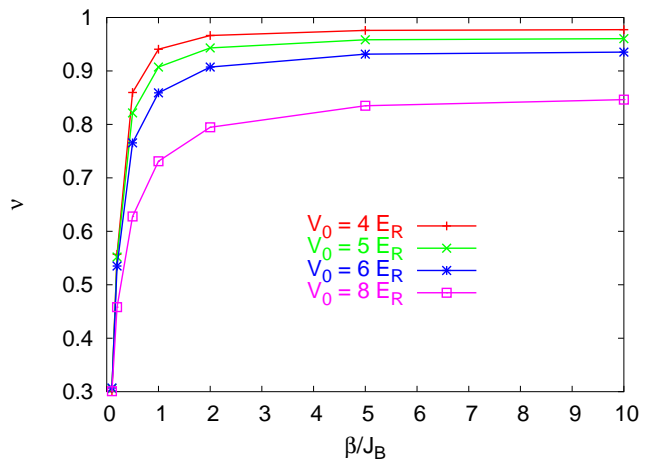}}
\caption{(Color online) Change in bosonic visibility when inverse temperature $\beta$ is increased for a system of 100 sites, 50 bosons and 30 fermions. }
\label{fig:temperature}
\end{figure}

\subsection{Comparison with experiment}\label{subsec:exp}

In apparent contradiction to our low-temperature predictions of an \emph{increase} in
 the bosonic visibility for most numbers of admixed fermions (in Fig.~\ref{fig:concentration}),
  the experiments~\cite{Gunter06, Ospelkaus06} show 
 a strong \emph{decrease}  for moderate
 values of the lattice potential. 
 Assuming that entropy is conserved when ramping up the lattice, the effective temperature of a  ${}^{87}$Rb -${}^{40}$K mixture rises dramatically because of the different temperature
 dependence of fermionic ($S_F\propto T/T_F$ for an ideal gas) and bosonic
 ($S_B\propto( T/T_c)^3$ for an ideal gas) contribution to the entropy~\cite{Gunter06}.
    In Fig.~\ref{fig:temperature}, we increase the temperature (decrease $\beta=1/k_BT$) at fixed optical
 potential and atom number and find a drastic drop in the visibility
 as the temperature is increased around $\beta \approx 2/ J_{\textrm{B}}$. 
A rise in the temperature of the bosonic atoms in the presence of fermions
can cause a large decrease of the bosonic visibility and is thus the most likely explanation for the experimental results.
% However, at deeper lattices temperature is probably so high that a single band Bose-Fermi Hubbard % model breaks down and that the gases behave essentially classically.

%%%%%%%%%%%%%%%%%%%%%%%%%%\paragraph{charge modulation}

\section{At low densities}
 
\begin{figure}[ t]
%\centerline{\includegraphics[scale=0.35,angle=270]{fig_polaron}}
\centerline{\includegraphics[angle=270, scale=0.35]{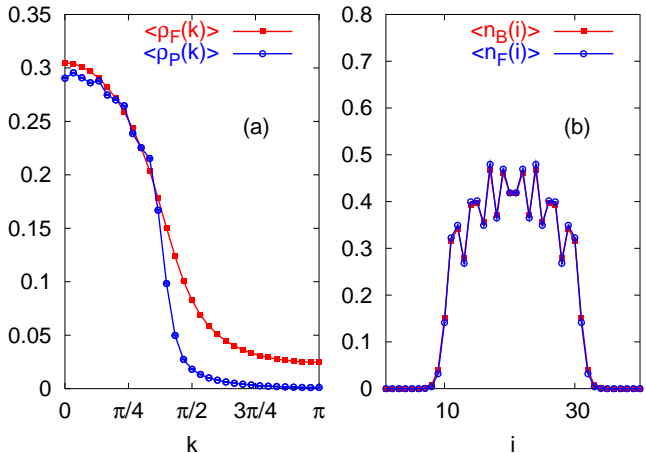}}
\caption{(Color online)  (a) Fermionic $\langle \rho_{\rm F}(k) \rangle$ and molecular Bose-Fermi  $\langle \rho_P(k) \rangle $ momentum profiles for the same system. 
   Error bars are shown, but are
  smaller than the point size. 
  (b) Bosonic and fermionic ground state density
  profiles for a system with parameters $L=60$,$N_{\rm B} = N_{\rm F} = 8, V_0 =
  4E_{\rm R} (U_{\rm BB} / J_{\rm B} = 6.13)$ determined using DMRG (QMC results agree within error bars). The bosonic and fermionic density profiles almost coincide. }\label{fig:polaron}
\end{figure}
 
%We turn now to the regime 
We finally discuss the physics at  low densities. 
%In this regime it turns out
%that considering the bosons in the mixture as `influenced' by the fermions is not always a
%suitable viewpoint. 
% In particular bound pairs (``molecules'') of one boson and one fermion \emph{can}
Bound pairs (``molecules'') of one boson and one fermion \emph{can} now
be formed for moderately deep optical lattices. First signs of this pairing can be seen in the
two-body Bose-Fermi momentum distribution shown in Fig.~\ref{fig:polaron} (a). The
momentum distribution of these fermionic molecules shows a sharper Fermi edge compared
to the bare fermion. It also tends to zero at larger momenta, showing that these molecules are a better description of the system than the bare fermions and bosons.
The formation of molecules is also well supported by coinciding charge
modulations in the bosonic and fermionic densities (Fig.~\ref{fig:polaron}(b)).
For the parameters chosen in Fig.~\ref{fig:polaron}(b) the density modulations are Friedel oscillations due to trap, but for larger lattice depths a density wave can be formed \cite{Pollet06,  Mathey04}.

\section{Conclusion}

In conclusion, we have simulated the trapped one-dimensional Bose-Fermi
Hubbard model. The interplay between temperature, trap, optical potential and
particle number is very rich and non-universal. The dominant effect is the creation of a strongly inhomogeneous trapping potential by the fermions. 
In higher order the fermions induce an attractive interaction between the
bosons, which should lead to an increase in the bosonic visibility. However
  assuming a rise in temperature when ramping up the lattice, a decrease in
  the visibility is found analogous to the experimental observation in the ${}^{87}$Rb -${}^{40}$K
  samples. If temperature remains low, one could observe a Mott transition driven by the fermionic concentration, and observe the formation of molecules.  The same effects are expected for higher dimensions, since the underlying mechanisms do not depend on dimensionality.
  %: the modification of the effective trapping potential and the fermion-induced attraction between the bosons willoccur in any dimension, as is readily seen in perturbation theory. 
  %The experimentally observed visibility depends on the formation or destruction of Mott regions and on the temperature -- again both effects are qualitatively the same in higher dimensions.
%With this study of the Bose-Fermi system we show that in all realizations of
%cold atom gases up to know it is neccessary to analyse
%carefully the influence of the temperature, trap, optical potential, since
%these may give rise to new physical effects and obscure the physics expected
% in homogeneous systems.
Our study explicitly shows that the effects of temperature, particle number, adiabatic processes and trapping potential have to be taken into account carefully when analyzing cold-atom experiments.

We are grateful to H.P.~B{\"u}chler, T.L.~Ho, A.~Muramatsu, G.~Pupillo, B.V.~Svistunov and the group of
T.~Esslinger for stimulating discussions. We acknowledge support by the Swiss National
Science Foundation and the Aspen Center for Physics. Simulations were performed on the Hreidar Beowulf cluster at ETH Zurich.

\end{document}